%% file: two_cylinder.tex
\documentclass[aps,prb,amsmath,amssymb,twocolumn,floatfix,showpacs,superscriptaddress]{revtex4-1}

\usepackage{graphicx}
\usepackage[usenames,dvipsnames,svgnames,table]{xcolor}
\usepackage{hyperref}
\hypersetup{
	colorlinks=true,       
	linkcolor=VioletRed,  
    citecolor=JungleGreen,        
    filecolor=Orchid,      
    urlcolor=JungleGreen           
}

\usepackage{hhline} 
\usepackage{epstopdf}
\usepackage{times}
\usepackage{multirow}

\DeclareMathOperator{\arccot}{arccot}

\begin{document}

\title{Shape dependence of two-cylinder R\'{e}nyi entropies for free bosons on a lattice}

\author{Leilee Chojnacki}
\affiliation{Perimeter Institute for Theoretical Physics, Waterloo, Ontario, N2L 2Y5, Canada}

\author{Caleb Q. Cook}
\affiliation{Perimeter Institute for Theoretical Physics, Waterloo, Ontario, N2L 2Y5, Canada}

\author{Denis Dalidovich}
\affiliation{Perimeter Institute for Theoretical Physics, Waterloo, Ontario, N2L 2Y5, Canada}

\author{Lauren E. Hayward Sierens}
\email{lhayward@perimeterinstitute.ca}
\affiliation{Perimeter Institute for Theoretical Physics, Waterloo, Ontario, N2L 2Y5, Canada}
\affiliation{Department of Physics and Astronomy, University of Waterloo, Ontario, N2L 3G1, Canada}

\author{\'{E}tienne Lantagne-Hurtubise}
\affiliation{Perimeter Institute for Theoretical Physics, Waterloo, Ontario, N2L 2Y5, Canada}

\author{Roger G. Melko}
\affiliation{Perimeter Institute for Theoretical Physics, Waterloo, Ontario, N2L 2Y5, Canada}
\affiliation{Department of Physics and Astronomy, University of Waterloo, Ontario, N2L 3G1, Canada}

\author{Tiffany J. Vlaar}
\affiliation{Perimeter Institute for Theoretical Physics, Waterloo, Ontario, N2L 2Y5, Canada}

\date{\today}
\pacs{} 

\begin{abstract}

Universal scaling terms occurring in R\'{e}nyi entanglement entropies have the potential to bring new 
understanding to quantum critical points in free and interacting systems.
Quantitative comparisons between analytical continuum theories and numerical calculations on lattice models play a crucial role in advancing such studies.
In this paper, we exactly calculate the universal two-cylinder shape dependence of entanglement entropies for free bosons on finite-size
square lattices, and compare to approximate functions derived in the continuum using several different ansatzes.
Although none of these ansatzes are exact in the thermodynamic limit, we find that numerical fits are in good agreement with 
continuum functions derived using the AdS/CFT correspondence, 
an extensive mutual information model, and a quantum Lifshitz model.
We use fits of our lattice data to these functions to calculate universal scalars defined in the thin-cylinder limit, and 
compare to values previously obtained for the free boson field theory in the continuum.

\end{abstract}

\maketitle

%
%

\section{Introduction}

Complementing traditional quantities used to understand critical phenomena such as scaling exponents,\cite{Stanley:1999} entanglement entropies have
begun to provide physicists with a wealth of new quantities that exhibit universality across a wide variety of physical theories.  
Defined through a geometrical bipartition of a system into two parts $A$ and its complement $\overline{A}$,
the entanglement entropy (and its generalized R\'{e}nyi entropies) obey novel scaling behaviour as the size of $A$ is varied.
The most famous example of a universal quantity extracted from such scaling is the central charge of a $(1+1)$-dimensional conformal
field theory (CFT), which can be obtained by varying the length of $A$.\cite{Holzhey:1994,Vidal:2003aa,Korepin:2004aa,Calabrese:2004}
In $d+1$ dimensions (for $d>1$) spatial geometries become highly non-trivial, and
 the entropy scaling is dominated by the ubiquitous ``area'' law, growing proportional to the boundary length 
between $A$ and $\overline{A}$.\cite{Bombelli1986,Srednicki1993}
The most useful universal numbers occur in scaling terms that are sub-leading to the area law.
Like the central charge in $d=1$, these numbers can potentially give deep insight into the low-energy theories governing
critical behaviour, providing for example an effective measure of degrees of freedom in the CFT\cite{Kallin:2014,Bueno:2015,Miao_2015,BuenoMyers2015,Faulkner:2015aa} or bounds on 
renormalization group flows,\cite{Zamolodchikov:1986,Atheorem,CHM_2011,Klebanov:2012,Grover:2014e} depending on the geometry of $A$.
Furthermore, they can be calculated in a host of models and theories, such as non-interacting lattice models, interacting Hamiltonians tuned to
a critical point, continuum CFTs, and strongly-interacting gravity duals through the AdS/CFT,\cite{RyuT}
providing non-trivial insight into the correspondences between these different universal theories.

Just as critical exponents, particularly for interacting systems, often rely on lattice calculations such as series expansions or Monte Carlo
for determination of their numerical values, so can lattice calculations provide access to universal numbers derived from entanglement entropy scaling.\cite{Singh:2012t}
This approach has been important for interacting Hamiltonians. 
For example, quantum Monte Carlo (QMC) is able to access the integer
$\alpha$ R\'{e}nyi entropies on finite-size tori.\cite{Hastings:2010,Tommaso,Inglis_2013,Helmes:2014,Helmes:2015}  
Numerical linked cluster expansions (NLCE), combined with exact diagonalization\cite{Kallin:2013} or the 
density-matrix renormalization group (DMRG),\cite{Kallin:2014,Miles:2014} can provide finite-cluster-sized estimates for all $\alpha$ entropies in the thermodynamic limit.
Lattice calculations have also proven important for accessing certain entangling geometries in free theories.\cite{Casini:2007,Casini:2009,Sahoo,Witczak_2016, Helmes:2016}

\begin{figure}[t]
    \def\svgwidth{0.5\columnwidth}
    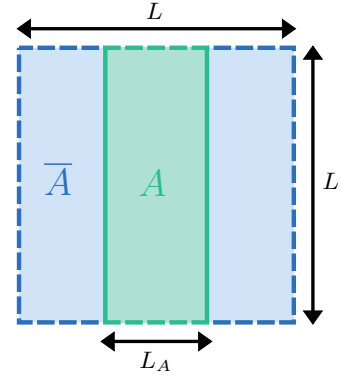
    \caption{An $L \times L$ square lattice divided in region $A$ and its complement, forming two cylinders.  Each outer (dashed) boundary may be periodic or anti-periodic. }
    \label{fig:torus}
\end{figure}

In this paper, we focus on one specific geometry in $2+1$ dimensions: a torus cut into two cylinders.  This is perhaps the most
straightforward entanglement bipartition possible on a finite-size lattice, in that it avoids sharp corners (which induce a sub-leading logarithmic term),\cite{Casini:2007,Bueno:2015,Bueno:2015JHEP,Bueno:2016} as well as curvatures in the boundary (which induce other sub-leading terms).\cite{Casini:2012,Liu2013,Casini2015,QLMcircle}
In $d=2$, with a two-cylinder geometry defined as in Fig.~\ref{fig:torus}, the R\'{e}nyi entropies are expected to scale with the region $A$ as
\begin{equation}
S_\alpha = a_\alpha \frac{L}{\delta} +  \gamma_\alpha(u) + \cdots
\label{eq:entropyScaling}
\end{equation}
where $\delta$ is the lattice cutoff, $u = L_A/ L$, $\alpha$ is the R\'{e}nyi index (to be defined in Sec.~\ref{sec:free_bosons}), and the ellipsis indicates non-universal constants and subleading terms depending on $\delta/L$ to some power.
The function $\gamma(u)$ is not analytically known in closed form even for the simplest non-interacting theories, but it is expected to reflect universality.\cite{Witczak_2016}
For a general lattice with linear sizes $L_x$ and $L_y$ respectively along the $x$- and $y$-directions, $\gamma(u)$ also depends upon the aspect ratio\cite{Witczak_2016,Fradkin_2015} $L_x/L_y$, but here we focus our attention on the case where $L_x=L_y=L$. 
In the thin-cylinder limit where $u \to 0$, this leading correction $\gamma(u)$ is expected to behave such that\cite{Casini:2009, Witczak_2016}
\begin{equation}
 \gamma_\alpha(u\to 0) = -\frac{\kappa_\alpha}{u},
\label{eq:thinCylinder}
\end{equation}
where $\kappa_\alpha$ is a universal constant.

Here, we numerically calculate $\gamma_\alpha(u)$ for all values of $u$ for non-interacting bosons on square lattices, and compare our results to several candidate functions derived from different ansatzes: the $(1+1)$-dimensional CFT scaling function,\cite{Holzhey:1994,Vidal:2003aa,Korepin:2004aa,Calabrese:2004} a quantum Lifshitz model (QLM),\cite{Stephan_2013}
anti de-Sitter (AdS) gravity in $3+1$ dimensions,\cite{Fradkin_2015} and an extensive Mutual Information (EMI) model.\cite{Casini_2005,Disconn,Swingle_2010,Bueno:2015JHEP}
Using lattices of different size, we obtain data for the entanglement entropies $S_{\alpha}$ for free bosons, 
and explore the finite-size scaling behaviour of the residuals between the data and these candidate functions.
We focus in particular on the von Neumann entropy $S_1$ and the second R\'{e}nyi entropy $S_2$, though we find similar trends for other R\'{e}nyi indices $\alpha>0$.
Finally, we calculate the universal numbers $\kappa_1$ and $\kappa_2$ from fits of our numerical data to the QLM, AdS and EMI functions, and compare to 
values obtained from continuum\cite{Casini:2009,Bueno:2015JHEP} and lattice calculations in the thin-cylinder limit.

\section{Two-cylinder scaling functions}
\label{sec:scalingFuncs}

Attempts to understand the quantum critical R\'{e}nyi entropy scaling of Fig.~\ref{fig:torus} through finite-size lattice numerics began with QMC simulations on interacting systems in $2+1$ dimensions.  
For a critical resonating-valence bond (RVB) wavefunction on a square lattice, Ju {\it et al.}\cite{Ju_2012}~postulated heuristically that the well-known $(1+1)$-dimensional CFT
scaling function,
\begin{equation}
\gamma_{\rm \scriptscriptstyle 1+1}(u) = c \, \ln  \sin (\pi u),
\label{eq:gamma_1D}
\end{equation}
could apply for fixed $L$ (where $c$ is a constant that is proportional to the central change in $1+1$ dimensions).  
Subsequent examination on several interacting and non-interacting systems in $2+1$ dimensions show that, although 
this function works approximately, quantitative differences remain between it and finite-size lattice data extrapolated to
the thermodynamic limit.\cite{Ju_2012,Inglis_2013,Fradkin_2015}
For instance, $\gamma_{\rm \scriptscriptstyle 1+1}(u)$ does not obey Eq.~\eqref{eq:thinCylinder} in the thin-cylinder limit.

An improvement on this form, motivated by the study of dimer RVB wavefunctions in the continuum limit, was derived
by St\'{e}phan {\it et al.} for the quantum Lifshitz model\cite{Stephan_2013} (QLM) with dynamical scaling exponent $z=2$. It is given by
\begin{equation}
\gamma_{\rm \scriptscriptstyle QLM}(u) 
= \frac{24\,\kappa}{\pi} \,\ln \!\left( \frac{\eta(2 i u)\,\eta(2 i (1-u))}{\theta_3(i \lambda u )\,\theta_3(i \lambda(1-u))}\right),
\label{eq:gamma_QLM}
\end{equation}
sometimes called $J(u)$ in the literature.  Here, $\theta_3$ is the Jacobi theta-function, $\eta$ is the Dedekind eta-function,
and $\lambda$ is a model-dependent parameter.
In the following, we fix $\lambda=2$ as for the dimer model in Ref.~\onlinecite{Stephan_2013},
although in principle this parameter could have a different value.
Comparison of this function to finite-size scaling data on conformally-invariant ($z=1$) critical points shows a surprisingly 
accurate fit, leading to the early speculation that it could be a universal 
scaling function relevant for {\it all} fixed points in $2+1$ dimensions, not just those specific to the QLM.\cite{Inglis_2013,Fradkin_2015}

X.~Chen {\it et al.}~invoked the AdS/CFT correspondence to propose another candidate function, derived in $3+1$ dimensions using the AdS soliton metric.\cite{Fradkin_2015}  This holographic function is in a parameterized form; up to a constant, it is (for an $L \times L$ lattice),
\begin{equation}
\gamma_{\rm \scriptscriptstyle AdS}(\chi) = \frac{\kappa \: \Gamma^4\!\left( \frac{1}{4}  \right)}{3\pi^2} \,\chi^{-1/3} \!\left(   \int_0^1 \frac{ d\zeta}{\zeta^2} \left( \frac{1}{ \sqrt{P(\chi,\zeta)}}  -1  \right)  \! -1 \right)
\label{eq:gamma_AdS}
\end{equation}
where $P(\chi,\zeta) = 1- \chi \zeta^3 - (1-\chi) \zeta^4$ and $\Gamma$ is the gamma function.  The parameter $\chi$ is related to the aspect ratio $u=L_A/L$ through the equation
\begin{equation}
u (\chi) = \frac{3 \chi^{1/3} (1-\chi)^{1/2}}{2 \pi} \int_0^1 \frac{ d \zeta \zeta^2}{ (1- \chi \zeta^3)} \frac{1}
{ \sqrt{P(\chi,\zeta)} } .
\end{equation}
Numerically, this function appears to describe very well the subleading entropy scaling term of two scale-invariant fermionic models: free massless Dirac fermions and a model of fermions with quadratic band touching.\cite{Fradkin_2015}

Most recently, Witczak-Krempa {\it et al.}~used an extensive mutual information (EMI) model,\cite{Casini_2005,Disconn,Swingle_2010,Bueno:2015JHEP} to derive another functional form,\cite{Witczak_2016}
\begin{equation}
\gamma_{\rm \scriptscriptstyle EMI}(u) = \frac{2\kappa}{\pi}\left[ \frac{\arccot \left( 2u \right) }{u} + \frac{\arccot \left( 2(1-u)  \right) }{1-u} \right].
\label{eq:gamma_EMI}
\end{equation}
The EMI model has been shown to be useful in the analysis of entropy scaling in CFTs\cite{Bueno:2015,Bueno:2015JHEP,Witczak_2016}.  

In the next section, we introduce our calculation for the free boson field theory on a finite-size lattice, and use numerical
solutions for the R\'{e}nyi entropies to evaluate each of the candidate scaling functions outlined above.

\section{Free bosons on the square lattice}
\label{sec:free_bosons}

Beginning with the action for a free real scalar (Klein-Gordon field) $\phi$ of mass $m$ in $d+1$ dimensions,
one can regularize the theory on a finite two-dimensional square lattice, such that the field and its conjugate momentum $\pi_i$ exist at each lattice site $i$ and evolve according to the Hamiltonian
\begin{align}\label{eq:bosonhamiltonian}
H = \frac{1}{2} \sum _{x,y =1,1}^{L_{x},L_{y}} \Big[ \pi_{x,y}^{2} &+  (\phi_{x+1,y} - \phi_{x,y})^{2} +
 (\phi_{x,y+1} - \phi_{x,y})^{2}\nonumber\\ +& \;m^{2} \phi _{x,y} ^{2}\Big].
\end{align}
Here, $L_x$ and $L_y$ are the linear dimensions of the lattice, and the total number of sites is $N=L_x \times L_y$. 
Transforming into Fourier space, the Hamiltonian can be written in the form of $N$ uncoupled simple harmonic oscillators,
\begin{equation} \label{eq:Hamiltonian_SHM}
H =  \frac{1}{2} \sum_\mathbf{k} \left[ \pi_\mathbf{k} \pi_\mathbf{-k} + \omega_\mathbf{k}^2 \phi_\mathbf{k} \phi_\mathbf{-k} \right],
\end{equation}
where
\begin{equation}
\omega_\mathbf{k} = \sqrt{4 \sin^2 \left( k_x /2 \right) + 4 \sin^2 \left( k_y /2 \right) + m^2}.
\end{equation}
The ground-state two-point correlation functions are given in Fourier space by
\begin{align}
\left\langle \phi_\mathbf{k} \phi_\mathbf{-k'} \right\rangle &= \frac{1}{2\omega_\mathbf{k}} \delta_{\mathbf{k}\mathbf{k'}}, \\
\left\langle \pi_\mathbf{k} \pi_\mathbf{-k'} \right\rangle 
&= \frac{\omega_\mathbf{k}}{2} \delta_{\mathbf{k}\mathbf{k'}}.
\end{align}

These correlation functions can be transformed back to real space, and by restricting the numbers of lattice points, the momenta are quantized. 
On a translationally invariant lattice, we thus obtain
\begin{align}
\left\langle \phi_\mathbf{x} \phi_\mathbf{x'} \right\rangle &= 
\frac{1}{2L_x L_y } \sum_\mathbf{k}  
\frac{\cos \left[ k_x\left( x - x'\right)\right] \cos \left[ k_y\left( y - y'\right)\right] } {\omega_\mathbf{k}} \nonumber \\
\left\langle \pi_\mathbf{x} \pi_\mathbf{x'} \right\rangle &= \frac{1}{2L_x L_y } \sum_\mathbf{k} 
\omega_\mathbf{k}\cos \left[ k_x\left( x - x'\right)\right] \cos \left[ k_y\left( y - y'\right)\right] .
\end{align}
If one considers periodic boundary conditions (PBC) for the fields, the momentum sums are restricted to
$k_i = 2 n_i \pi/L_i$, where $n_i = 0, 1, \ldots, L_i -1$.  
Note that the case where the boson is massless, $m=0$, is desired in order to obtain a scale-invariant critical theory, but the presence of the zero-mode $k_x = k_y = 0$ for PBC in this case causes the $\phi$ correlation function to diverge.
The entanglement entropy thus cannot be calculated directly for a fully periodic system in the massless (critical) case, and for such systems we include a small but finite mass of $m = 10^{-6}$.
Alternatively, in many of the calculations below, we will employ an anti-periodic boundary condition (APBC), e.g.~$\phi_{x} = -\phi_{x + L_x}$, in at least one lattice direction, which gives
$k_i = (2 n_i + 1) \pi/L_i$ and avoids this divergence (even when $m=0$).

We employ these correlation functions to calculate the von Neumann entanglement entropy, defined in terms of the reduced density matrix $\rho_A = {\rm Tr}_B \left( \rho_{AB} \right) $ as
\begin{equation} \label{eq:vonNeumann_EE}
S_1(A) = -{\rm Tr} \left( \rho_A \log \rho_A \right).
\end{equation}
We also study the generalized R\'{e}nyi entanglement entropies, which are given by
\begin{equation} \label{eq:Renyi_EE}
S_\alpha (A) = \frac{1}{1-\alpha} \log \left( {\rm Tr} \rho_A^\alpha \right),
\end{equation}
where $\alpha$ is called the R\'{e}nyi index.  The case of $\alpha=2$ is particularly important for QMC studies of strongly-interacting lattice Hamiltonians.\cite{Hastings:2010}
Note that taking the limit $\alpha \to 1$ in Eq.~\eqref{eq:Renyi_EE} recovers the (von Neumann) entanglement entropy of Eq.~\eqref{eq:vonNeumann_EE}. 

As discussed by Peschel in Ref.~\onlinecite{Peschel} and by Casini and Huerta in Ref.~\onlinecite{Casini:2009}, 
the reduced density matrix for a non-interacting system such as this one can be written as
$\rho_A = \mathcal{K} e^{-H_A},$
where $\mathcal{K}$ is a normalization constant and $H_A$ is called the \textit{modular Hamiltonian}, which is quadratic and acts only on sites in region $A$.  
The von Neumann and R\'{e}nyi entropies can be calculated from  
the correlation functions within region $A$, thus avoiding the trace over $\overline{A}$ in the calculation of the reduced density matrix. 
Defining the elements of our \textit{correlation matrices} $X_A$ and $P_A$ such that
$\left( X_A \right)_{ij} = \left\langle \phi_\mathbf{x_i} \phi_\mathbf{x_j} \right\rangle$
and 
$\left( P_A \right)_{ij} = \left\langle \pi_\mathbf{x_i} \pi_\mathbf{x_j} \right\rangle$
for sites $\mathbf{x_i}$ and $\mathbf{x_j}$ in region $A$,
the entropies are then given in terms of the eigenvalues $\nu_\ell$ of the matrix $C_A = \sqrt{X_A P_A}$ by
\begin{align}
S_1 (A) = 
\sum_\ell \bigg[ & \left( \nu_\ell + \frac{1}{2} \right) \log \left( \nu_\ell + \frac{1}{2} \right) \\ \nonumber
&- \left( \nu_\ell - \frac{1}{2} \right) \log \left( \nu_\ell - \frac{1}{2} \right) \bigg], \label{eq:vonNeumann_EE_eigvals}
\end{align}
and
\begin{equation}
S_\alpha(A) = 
\frac{1}{\alpha-1} \sum_\ell 
\log \left[ \left( \nu_\ell + \frac{1}{2} \right)^\alpha - \left( \nu_\ell - \frac{1}{2} \right)^\alpha \right].\label{eq:Renyi_EE_eigvals}
\end{equation}

The procedure described above allows us to calculate the two-cylinder entropy $S_\alpha(u)$ for the full range of $u$ values for lattice sizes up to $L=240$ using modest computational resources.
In addition, for the two-cylinder geometry discussed in this paper, we employ an extension of the above arguments as given in Ref.~\onlinecite{Fradkin_2015} that takes advantage of the translational symmetry in one lattice direction to map the 
$(2+1)$-dimensional model to an effective model consisting of $L$ separate $(1+1)$-dimensional chains.
This mapping allows for the calculation of R\'{e}nyi entropies on significantly larger lattices.

\section{Results}

\begin{figure}[t]
    \includegraphics{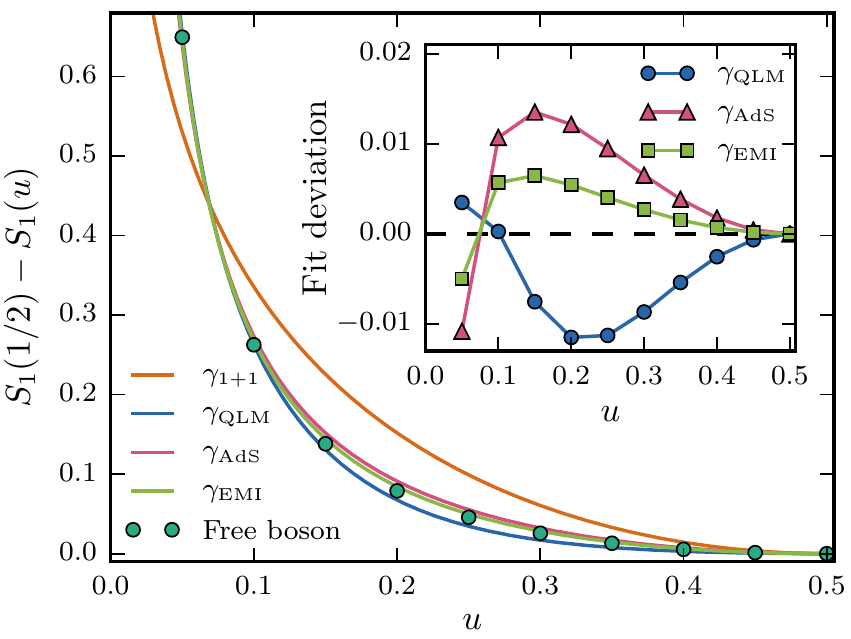}
    \caption{ Shape dependence of the von Neumann entanglement entropy for an $L=3000$ free boson system with PBC along the $x$-direction and APBC along the $y$-direction, and fits corresponding to the four candidate functions discussed in Sec.~\ref{sec:scalingFuncs} (each with resolution $\Delta u =0.05$). The inset illustrates the amount that each fit deviates from each data point.  The fit deviation for $\gamma_{\rm \scriptscriptstyle 1+1}$ is excluded since it is much larger in magnitude than the other three curves. }
    \label{fig:vNFits}
\end{figure}

We use the procedure of the last section to calculate the von Neumann ($S_1$) and second R\'{e}nyi ($S_2$) entropies
for lattice geometries of size $L \times L$, with $L_A$ varying from 1 to $L/2$ (see Fig.~\ref{fig:torus}).
An example of a finite-size lattice calculation of $S_1$ as a function of $u$, for a fixed value of $L$, is illustrated in Fig.~\ref{fig:vNFits}.

For fixed $L$, the area law and subleading terms in Eq.~\eqref{eq:entropyScaling} are constant such that, as a function of $u$, we have $S_\alpha(u) = \gamma_\alpha(u) + d_\alpha$, where $d_\alpha$ is a constant ($L$-dependent) parameter and the function $\gamma_\alpha(u)$ contains the additional parameter $\kappa_\alpha$ (or $c_\alpha$ for $\gamma_{\rm \scriptscriptstyle 1+1}$).
In order to reduce the number of fitting parameters from two to one, we perform least-squares fits of $S_\alpha(1/2) - S_\alpha(u)$ to the form $\gamma_\alpha(1/2) - \gamma_\alpha(u)$ such that $\kappa_\alpha$ (or $c_\alpha$) becomes the sole fitting parameter. 
Fig.~\ref{fig:vNFits} illustrates such fits for the four candidate functions for $\gamma_\alpha(u)$ from Eqs.~\eqref{eq:gamma_1D}, \eqref{eq:gamma_QLM}, \eqref{eq:gamma_AdS} and \eqref{eq:gamma_EMI} for the case where $\alpha=1$.
From this, it is immediately obvious that $\gamma_{\rm \scriptscriptstyle 1+1}$ provides a poor approximation to the data.
The inset shows the amount that the three best fits deviate from the free boson as a function of $u$.

While this procedure provides the deviation of the free boson from the candidate functions for a given finite-size lattice, one
may ask how this deviation behaves as one scales the lattice size towards the thermodynamic limit.
Thus, we vary the system size and repeat the calculation for a range of values of $L$. 
In order to quantify the goodness of each one-parameter fit, we sum the squared residuals and normalize by the number of degrees of freedom,
defining the fitting error by
\begin{equation}
E_\alpha = \frac{1}{n_u \!- 2} \sum_{i=1}^{n_u} \! \bigg( \! \big[\gamma_\alpha\!\left(1/2\right) - \gamma_\alpha(u_i) \big] - \big[ S_\alpha\!\left(1/2\right) - S_\alpha(u_i) \big]  \! \bigg)^{\!\! 2}. \label{ErrorF}
\end{equation}
Here, the data points $S_\alpha(u_i)$ are the calculated free boson entanglement entropies.
The fitted function is $\gamma_\alpha(u_i)$, with fitting parameter $\kappa_\alpha$ (or $c_\alpha$ in $d=1$).
Finally, $n_u$ is the number of values of $u$ used within the fitting procedure.

In doing these fits, we find that the errors are especially sensitive to the data points at small $u$. 
In particular, if the fitting procedure uses all $L/2$ available data points, then the errors appear to diverge as the lattice size $L$ increases.
However, this divergence can be attributed to the fact that the resolution of a lattice scales according to $\Delta u = 1/L$, and thus larger lattices are capable of probing smaller values of $u$. 
Since these small-$u$ effects are not what we wish to measure, we perform our fits using a resolution $\Delta u$ and corresponding number of data points $n_u = 1/(2\Delta u)$ that remain fixed as the lattice size $L$ increases.
Such a constraint limits the lattice sizes on which we perform our fits to multiples of $1/\Delta u$.

\begin{figure}[t]
	\includegraphics{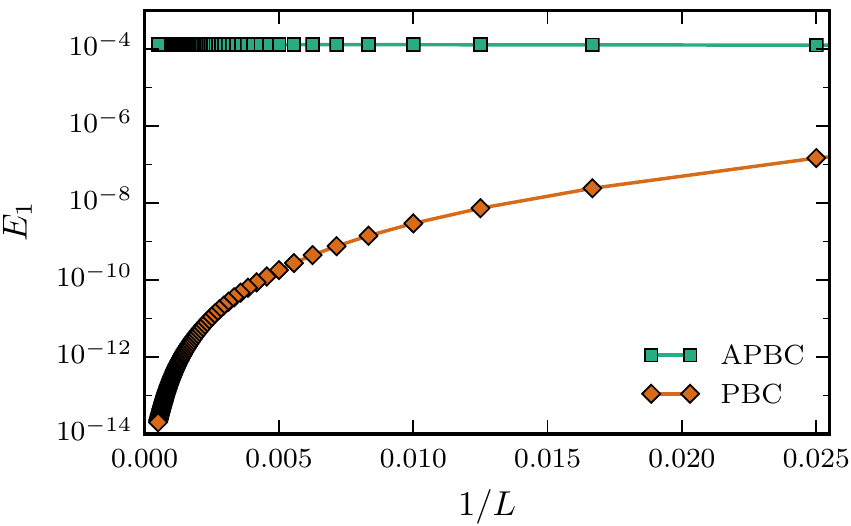}
	\caption{ The fitting errors corresponding to fits of the $(1+1)$-dimensional free boson von Neumann entropies to the expression in Eq.~\eqref{eq:gamma_1D}. We observe that this expression is exact in the thermodynamic limit (\textit{i.e.} the errors trend to zero) for PBC but not for APBC. These errors are measured using resolution $\Delta u =0.05$ on lattices of size $L = 40, 60, \ldots, 2000$. For the periodic case, we set $m=10^{-6}$ as explained in the text. }
	\label{fig:fitErrors_1D}
\end{figure}

One can employ this type of fixed-resolution fitting approach to evaluate a $(1+1)$-dimensional free boson system, which is known to 
exactly obey the scaling function in Eq.~\eqref{eq:gamma_1D} in the thermodynamic limit. 
As shown in Fig.~\ref{fig:fitErrors_1D}, we indeed find for free bosons in $1+1$ dimensions that the fitting errors corresponding to Eq.~\eqref{eq:gamma_1D} trend to zero (within machine precision) for PBC, validating our use of Eq.~\eqref{ErrorF} to quantify the residuals.
Interestingly, for APBC, the fitting error does not disappear in the limit $L \rightarrow \infty$, indicating that Eq.~\eqref{eq:gamma_1D} is not the 
correct description of the entanglement correction $\gamma_\alpha$ for this boundary condition.

\begin{figure*}[t]
	\includegraphics{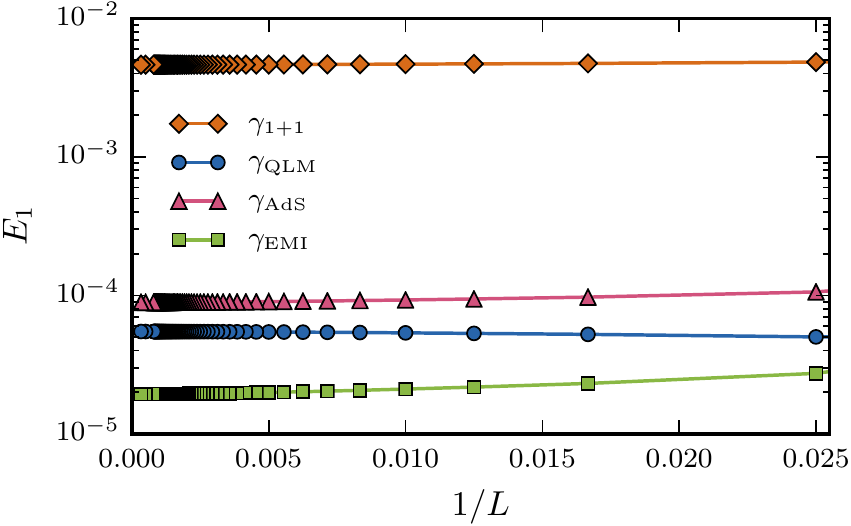}
	\hspace{1.5em}
	\includegraphics{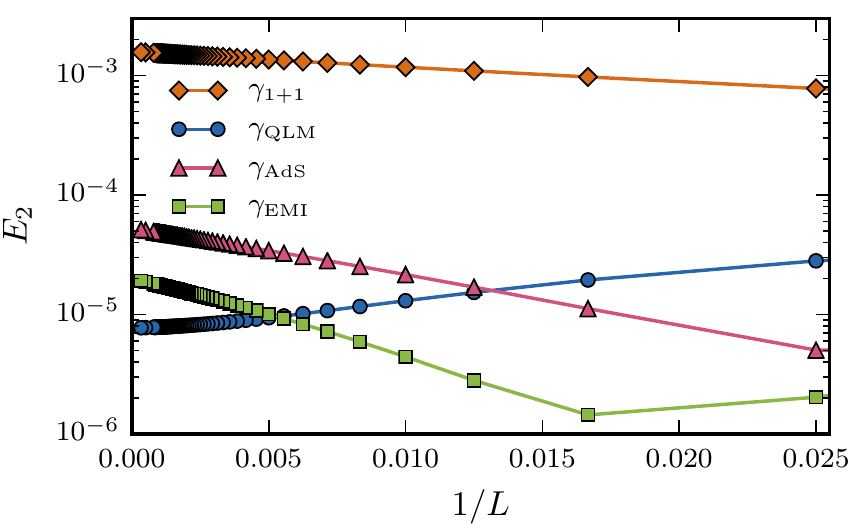}
	\caption{ The fitting errors corresponding to the four candidate functions for $\gamma$ as a function of $1/L$ for the 2D free boson von Neumann (left) and second R\'{e}nyi (right) entropies. 
	Both plots have PBC along the $x$ direction and APBC along the $y$ direction.
	The errors are measured using resolution $\Delta u =0.05$ on lattices of linear size $L = 40 \text{ up to } 3000$, as explained in the main text.
	}
	\label{fig:fitErrors}
\end{figure*}

\setlength{\tabcolsep}{0.8em}
\renewcommand{\arraystretch}{1.3}
\begin{table*}[t]
\begin{center}
\begin{tabular}{ | c | c | c | c | c | c | c | }
\hline
 Fitting error & $x$-direction & $y$-direction & 1+1 & QLM & AdS & EMI \\ 
\hhline{|=|==|====|}
           & PBC  & PBC & $3.18 \times 10^{-3}$ & $4.13 \times 10^{-3}$ & $1.56 \times 10^{-3}$ & $2.25 \times 10^{-3}$  \\ 
$E_1$ & PBC  & APBC & $4.63 \times 10^{-3}$   & $5.49 \times 10^{-5}$ & $8.90 \times 10^{-5}$ & $1.94 \times 10^{-5}$ \\  
           & APBC & PBC  & $4.60 \times 10^{-3}$ & $4.24 \times 10^{-5}$ & $1.12 \times 10^{-4}$ & $3.27 \times 10^{-5}$ \\  
           & APBC & APBC & $5.05 \times 10^{-3}$ & $2.85 \times 10^{-5}$ & $3.10 \times 10^{-4}$ & $1.69 \times 10^{-4}$ \\  
\hline
           & PBC  & PBC  & $9.28  \times 10^{-4}$ & $2.61 \times 10^{-3}$ & $1.19 \times 10^{-3}$ & $1.59 \times 10^{-3}$ \\
$E_2$ & PBC  & APBC & $1.56 \times 10^{-3}$ & $7.73 \times 10^{-6}$ & $5.12 \times 10^{-5}$ & $1.91 \times 10^{-5}$ \\ 
           & APBC & PBC  & $1.51 \times 10^{-3}$ & $9.94 \times 10^{-6}$ & $4.68 \times 10^{-5}$ & $1.70 \times 10^{-5}$ \\ 
           & APBC & APBC & $1.67 \times 10^{-3}$ & $2.06 \times 10^{-5}$ & $1.35 \times 10^{-4}$ & $8.28 \times 10^{-5}$ \\
\hline
\end{tabular}
\end{center}
	\caption{ 
	The fitting errors corresponding to the four candidate functions and different boundary conditions for the von Neumann and second R\'{e}nyi entropies. 
	The errors in this table are measured using resolution $\Delta u =0.05$ on a square lattice of size $L=3000$. 
	Calculations for the fully periodic system include a small mass $m=10^{-6}$. }
	\label{tab:fitErrors}
\end{table*}

\begin{figure*}[t]
    \includegraphics{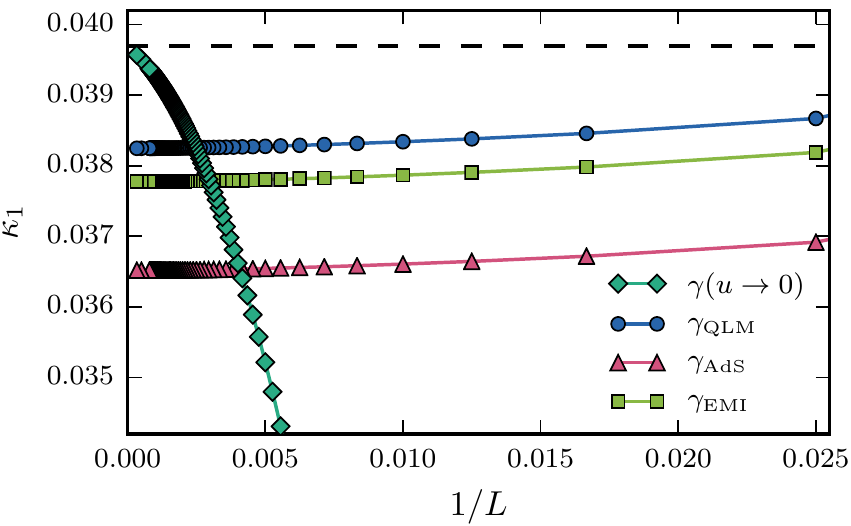}
    \hspace{1em}
    \includegraphics{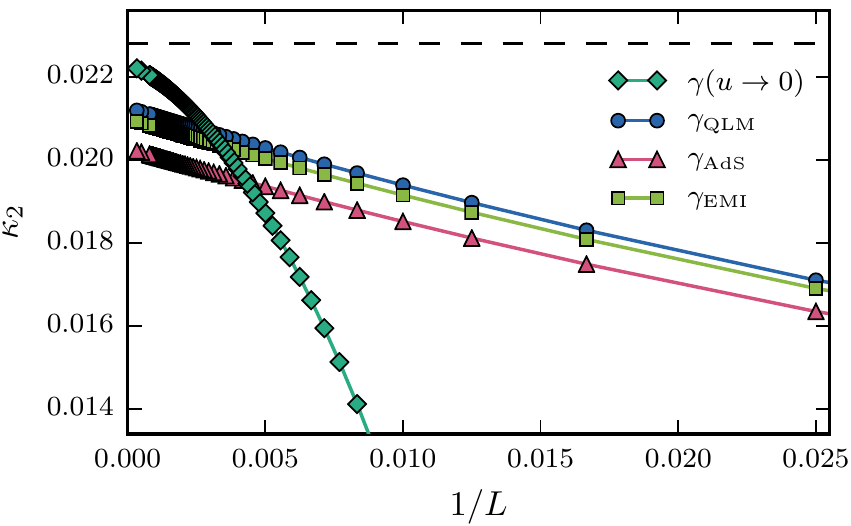}
    \caption{ The universal number $\kappa_\alpha$ for the von Neumann (left) and second R\'{e}nyi (right) entropies, as extracted from fits to $\gamma (u \to 0)$, $\gamma_{\rm \scriptscriptstyle QLM}$, $\gamma_{\rm \scriptscriptstyle AdS}$ and $\gamma_{\rm \scriptscriptstyle EMI}$ (as described in the main text). The dashed lines indicate the values calculated in the continuum in the $u \to 0$ limit\cite{Casini:2009,Bueno:2015JHEP}. }
    \label{fig:kappa}
\end{figure*}

In $2+1$ dimensions, the errors $E_\alpha$ have different finite-size scaling trends for the different candidate functional forms of $\gamma_\alpha$, as illustrated in Fig.~\ref{fig:fitErrors}.
We find that the behaviour of the error trends also depend both upon the chosen resolution $\Delta u$ and 
upon the lattice boundary conditions.
PBC along the $x$-direction and APBC along the $y$-direction are used 
to generate the plots presented in Figs.~\ref{fig:vNFits},~\ref{fig:fitErrors} and~\ref{fig:kappa}.
Table~\ref{tab:fitErrors} summarizes the fitting 
errors measured using resolution $\Delta u =0.05$ for various boundary conditions.
For the von Neumann entropy $S_1$, the functions $\gamma_{\rm \scriptscriptstyle QLM}$ and $\gamma_{\rm \scriptscriptstyle EMI}$ consistently yield the lowest fitting errors out of the four candidate functions, once at least one boundary is anti-periodic. 
It is interesting that for the case of periodic boundary conditions in both directions, the fitting errors for 
$\gamma_{\rm \scriptscriptstyle QLM}$,  $\gamma_{\rm \scriptscriptstyle EMI}$ and $\gamma_{\rm \scriptscriptstyle AdS}$ all become much larger, while for $\gamma_{\rm \scriptscriptstyle 1+1}$ these errors change only slightly and no longer correspond to the worst fit.
The explanation of this behaviour may lie in the effects of additional subleading scaling terms, which are induced by the presence of a finite mass $m$,\cite{Max_Tarun} on each of these fits.
In the case of the second R\'{e}nyi entropy, the errors corresponding to $\gamma_{\rm \scriptscriptstyle QLM}$  
are consistently lowest, except (again) for the case of PBC in both directions.

In addition to exploring the most suitable functional form for $\gamma_\alpha(u)$ for $\alpha = 1 \text{ and } 2$, we also examine the ability of our fits to extract universal numbers such as the coefficient $\kappa_\alpha$. 
For $(2+1)$-dimensional massless real free bosons in the continuum, this coefficient $\kappa_1$ has been calculated numerically for the von Neumann entropy\cite{Casini:2009} to be $\kappa_1=0.0397$ as well as for the second R\'{e}nyi entropy\cite{Bueno:2015JHEP} to be $\kappa_2 = 0.0227998$.
On a lattice, one can calculate $\kappa_\alpha$ by fitting to $\gamma(u \to 0)$ in Eq.~\eqref{eq:thinCylinder} for small $u$. 
Here we use a slightly different procedure than the fixed-resolution approach used to fit the four candidate functions:
for a given lattice of size $L \geq 80$, we extract $\kappa_\alpha$ in the $u \to 0$ limit from our free boson calculations by fitting $S_\alpha(40/L) - S_\alpha(u)$ to Eq.~\eqref{eq:thinCylinder} for $u = 31/L, 32/L, \ldots, 40/L$ (we ignore the smallest 30 values of $u$ due to numerical issues that arise when the cylinder becomes very thin).  Results are illustrated in Fig.~\ref{fig:kappa}.

In addition, the QLM, AdS and EMI functions all obey Eq.~\eqref{eq:thinCylinder} in the small-$u$ limit, allowing for predictions of the 
universal number $\kappa$ from each.
In Fig.~\ref{fig:kappa} we illustrate the $\kappa_1$ and $\kappa_2$ coefficients as predicted from fits (for the entire range of $u$ values and with fitting resolution $\Delta u=0.05$) to $\gamma_{\rm \scriptscriptstyle QLM}$, $\gamma_{\rm \scriptscriptstyle AdS}$ and $\gamma_{\rm \scriptscriptstyle EMI}$. 
Although we know from the fitting errors that none of these three candidate functions are exact in the thermodynamic limit, they are all still capable of extracting estimates for $\kappa_\alpha$ that agree relatively well 
with the previously-calculated continuum values\cite{Casini:2009,Bueno:2015JHEP} and the lattice values from fits to Eq.~\eqref{eq:thinCylinder}.  
In particular, $\gamma_{\rm \scriptscriptstyle QLM}$ and $\gamma_{\rm \scriptscriptstyle EMI}$ both yield estimates for $\kappa_1$ ($\kappa_2$) that are within less than 5\% (9\%) of the value calculated in Ref.~\onlinecite{Casini:2009} (Ref.~\onlinecite{Bueno:2015JHEP}).

\section{Discussion}

In this paper, we have studied the shape-dependence of the bipartite R\'{e}nyi entanglement entropies for a system of free bosons in $2+1$ dimensions.
Of particular interest is the universal term $\gamma_\alpha(u)$ that occurs sub-leading to the area law, which depends on the ratio $u$ of the
length of an entangled cylindrical region $A$ to the length of the entire system.
For this and other CFTs, an analytical expression for this universal function is unknown.  However, several candidate functions have been used in the
literature to approximate $\gamma_\alpha(u)$.  Performing exact but finite-size lattice calculations for free bosons on a $(2+1)$-dimensional square lattice,
we evaluate the quality of several of these candidate functions in the limit of large lattice size by examining a fitting error between the data
and each respective function.

We concentrate mainly on the von Neumann ($S_1$) and second R\'{e}nyi ($S_2$) entropies.  For each of these quantities, the candidate function
derived heuristically from the known $(1+1)$-dimensional CFT scaling form performs poorly and does not produce the correct thin-cylinder behaviour of Eq.~\eqref{eq:thinCylinder}, indicating that it should not be used to approximate $(2+1)$-dimensional CFT data in the two-cylinder geometry.  
Three other candidate functions with the correct behaviour in the $u\to 0$ limit were examined, derived from:
a quantum Lifshitz model (QLM), anti de-Sitter (AdS) gravity in $3+1$ dimensions, and an extensive Mutual Information (EMI) model.
All three of these give quantitatively better fits when compared to the $(1+1)$-dimensional CFT form.  Worst performing is the 
AdS function.  
The QLM and EMI provide the best fits.  
The success of the QLM is perhaps surprising:  
it has an additional parameter $\lambda$ that is unknown and has been fixed to an arbitrary value in our fits, and
it is derived in a non-conformally
invariant theory with dynamical exponent $z=2$.  
Despite there being no theoretical reason to believe it should apply to our boson CFT, which has $z=1$, it describes our data 
relatively well.
Finally, we note that the quantitative error in the fits is significantly affected by different combinations of periodic and anti-periodic 
boundary conditions, indicating that the exact CFT function $\gamma_\alpha(u)$, whatever it is, will also depend on the
phase angle by which the field $\phi$ is twisted at the boundary.

This work illustrates the care that must be taken when making comparisons of analytical functions, derived in the continuum limit, to
exact but finite-size entropy data obtained for lattice models.  
Not only are the residuals of fits to such functions affected by the finite size of the lattice itself,
but also on the range and position of data chosen for the comparison.  
Nevertheless, with sufficient care, the synergy between continuum 
theories and lattice numerics can bear fruit, such as we have demonstrated with the extraction of the universal coefficients $\kappa_\alpha$.  
The ability to calculate and compare such universal scalar quantities in the lattice and the continuum is of crucial importance 
in the continuing effort to use entanglement entropies as tools to characterize both free and strongly-interacting 
quantum critical points.

\section*{Acknowledgments}

We acknowledge crucial discussions with X.~Chen, P.~Fendley, E.~Fradkin, A.~Ludwig, M.~Metlitski, J.-M.~St\'ephen, G.~Vidal and W.~Witczak-Krempa.
We appreciate the hospitality of the organizers of the Perimeter Institute Winter School, where this project was initially conceived,
and the Aspen Center for Physics, where it was completed.
The simulations were performed on the computing facilities of SHARCNET. 
L.H.S. gratefully acknowledges funding from the Ontario Graduate Scholarship, and \'{E}. L.-H. is partially funded by FRQNT. 
Support was provided by NSERC, the Canada Research Chair program, the Ontario Ministry of Research and Innovation, and the Perimeter Institute for Theoretical Physics. 
Research at Perimeter Institute is supported by the Government of Canada through the Department of Innovation, Science and Economic Development Canada and by the Province of Ontario through the Ministry of Research, Innovation and Science.

%
%

\bibliographystyle{apsrev}
\bibliography{Biblo}

\end{document}

%% file: torus_2.pdf_tex
\begingroup%
  \makeatletter%
  \providecommand\color[2][]{%
    \errmessage{(Inkscape) Color is used for the text in Inkscape, but the package 'color.sty' is not loaded}%
    \renewcommand\color[2][]{}%
  }%
  \providecommand\transparent[1]{%
    \errmessage{(Inkscape) Transparency is used (non-zero) for the text in Inkscape, but the package 'transparent.sty' is not loaded}%
    \renewcommand\transparent[1]{}%
  }%
  \providecommand\rotatebox[2]{#2}%
  \ifx\svgwidth\undefined%
    \setlength{\unitlength}{931.63432245bp}%
    \ifx\svgscale\undefined%
      \relax%
    \else%
      \setlength{\unitlength}{\unitlength * \real{\svgscale}}%
    \fi%
  \else%
    \setlength{\unitlength}{\svgwidth}%
  \fi%
  \global\let\svgwidth\undefined%
  \global\let\svgscale\undefined%
  \makeatother%
  \begin{picture}(1,1.10855435)%
    \put(0,0){\includegraphics[width=\unitlength,page=1]{torus_2.pdf}}%
    \put(0.39793091,1.11669379){\color[rgb]{0,0,0}\makebox(0,0)[lt]{\begin{minipage}{0.37783065\unitlength}\raggedright $L$\end{minipage}}}%
    \put(0.94000798,0.59534342){\color[rgb]{0,0,0}\makebox(0,0)[lt]{\begin{minipage}{0.37783065\unitlength}\raggedright $L$\end{minipage}}}%
    \put(0,0){\includegraphics[width=\unitlength,page=2]{torus_2.pdf}}%
    \put(0.08404983,0.62749457){\color[rgb]{0.17254902,0.43529412,0.7372549}\makebox(0,0)[lt]{\begin{minipage}{0.418977\unitlength}\raggedright \Large{$\overline{A}$}\end{minipage}}}%
    \put(0,0){\includegraphics[width=\unitlength,page=3]{torus_2.pdf}}%
    \put(0.37760363,0.04233674){\color[rgb]{0,0,0}\makebox(0,0)[lt]{\begin{minipage}{0.37783065\unitlength}\raggedright $L_A$\end{minipage}}}%
    \put(0.37447794,0.6025921){\color[rgb]{0.17647059,0.7372549,0.56862745}\makebox(0,0)[lt]{\begin{minipage}{0.25439166\unitlength}\raggedright \Large{$A$}\end{minipage}}}%
    \put(0,0){\includegraphics[width=\unitlength,page=4]{torus_2.pdf}}%
  \end{picture}%
\endgroup%